\documentclass[a4paper,fleqn]{cas-dc}

\usepackage[numbers]{natbib}

\def\tsc#1{\csdef{#1}{\textsc{\lowercase{#1}}\xspace}}
\tsc{WGM}
\tsc{QE}

\begin{document}
\let\WriteBookmarks\relax
\def\floatpagepagefraction{1}
\def\textpagefraction{.001}
\let\printorcid\relax 

\shorttitle{Design and fabrication of ultrasound linear array transducer used in ultrasound endoscope}    

\shortauthors{Zhengbao Yang et al.}

\title[mode = title]{Design and fabrication of ultrasound linear array transducer used in ultrasound endoscope}

\author[1]{Yuan Zhang}

\author[1]{Mingtong Chen}

\author[1]{Zhengbao Yang}
\cormark[1] 
\ead{zbyang@hk.ust} 
\ead[URL]{https://yanglab.hkust.edu.hk/}

\address[1]{The Hong Kong University of Science and Technology
Hong Kong, SAR 999077, China}

\cortext[1]{Corresponding author} 

\begin{abstract}
This report details the successful construction of an ultrasound imaging platform and the design and fabrication of a novel ultrasound endoscope probe. The project's primary objective was to establish a functional system for acquiring and processing ultrasound signals, specifically targeting minimally invasive endoscopic applications. The ultrasound imaging platform was primarily designed and developed based on Texas Instruments (TI) Evaluation Modules (EVMs). It enables the transmission of 32-channel high-voltage signals and the reception of echo signals, with on-chip signal amplification and acquisition capabilities. Furthermore, the platform integrates a complete Time Gain Control (TGC) imaging path and a Continuous Wave Doppler (CWD) path. In conjunction with host computer software, it supports imaging with linear array, convex array, and phased array probes. Concurrently, a 64-element, 5MHz center frequency, phased array linear ultrasound endoscopic probe was designed, aiming for miniaturization and optimal imaging performance. The fabrication and assembly of its matching layer, backing layer, 2-2 piezoelectric composite material, and electrodes were completed.

\end{abstract}




\maketitle

\section{Introduction}

Ultrasound imaging is a vital medical diagnostic tool, offering non-invasive, real-time visualization of soft tissues and blood flow. Endoscopic ultrasound (EUS) enhances this by integrating ultrasound with endoscopy, providing high-resolution images of adjacent organs\cite{1}. EUS excels over conventional methods due to its close proximity to target tissues, enabling higher frequencies for superior spatial resolution. This is crucial for lesion characterization, tumor staging, and guiding minimally invasive procedures like fine-needle aspiration\cite{2}. Its ability to combine imaging and intervention makes EUS essential in gastroenterology and pulmonology.
Phased array transducers are key to EUS image quality, allowing electronic beam steering and focusing without mechanical movement, ideal for confined endoscopic spaces\cite{3,4}. Their performance benefits from piezoelectric composite materials (e.g., 2-2 composites), which improve acoustic impedance matching, electromechanical coupling, and bandwidth, leading to better sensitivity and resolution\cite{5,6}. Modern ultrasound platforms include sophisticated signal processing. Time Gain Control (TGC) ensures uniform image brightness by compensating for signal attenuation. Continuous Wave Doppler (CWD) continuously measures blood flow, vital for vascular assessment\cite{7}. Integrating these features forms the core of current ultrasound systems.
Despite these advancements, challenges remain in developing high-performance, miniaturized EUS systems. This project addresses key engineering problems in platform construction and endoscopic probe design. First, building a robust ultrasound imaging platform is challenging. It requires precise control over high-voltage signal transmission and sensitive echo reception. Integrating on-chip amplification, acquisition, TGC, and CWD paths demands careful hardware design and signal optimization. The platform must also support linear, convex, and phased array probes, balancing versatility, performance, and noise reduction within component constraints\cite{8}.

Second, designing and fabricating a miniaturized endoscopic probe is difficult. Endoscopic use requires extremely small dimensions without sacrificing imaging performance. A 64-element, 5MHz phased array linear transducer involves overcoming challenges like element spacing, crosstalk, acoustic impedance matching, and integrating sensitive piezoelectric composites in a tiny volume\cite{9,10}. Precise fabrication of matching, backing, composite, and electrode layers is critical for acoustic properties. Maintaining signal integrity over the long endoscopic cable, especially for a high-channel count array, adds complexity. Ultimately, seamless integration of the hardware platform with host software is vital for real-time image processing. This involves developing robust communication and efficient algorithms for data handling, beamforming, and image reconstruction. The overall engineering problem is to successfully combine a high-fidelity ultrasound platform with a cutting-edge, miniaturized endoscopic probe into a functional system for minimally invasive applications.

\section{Construction of ultrasound imaging platform}

The ultrasound imaging platform consists of the following parts: transmitting circuit, receiving circuit, digital beamformer, data communication, and power supply.

\begin{figure*}[h]
	\centering
		\includegraphics[scale=1]{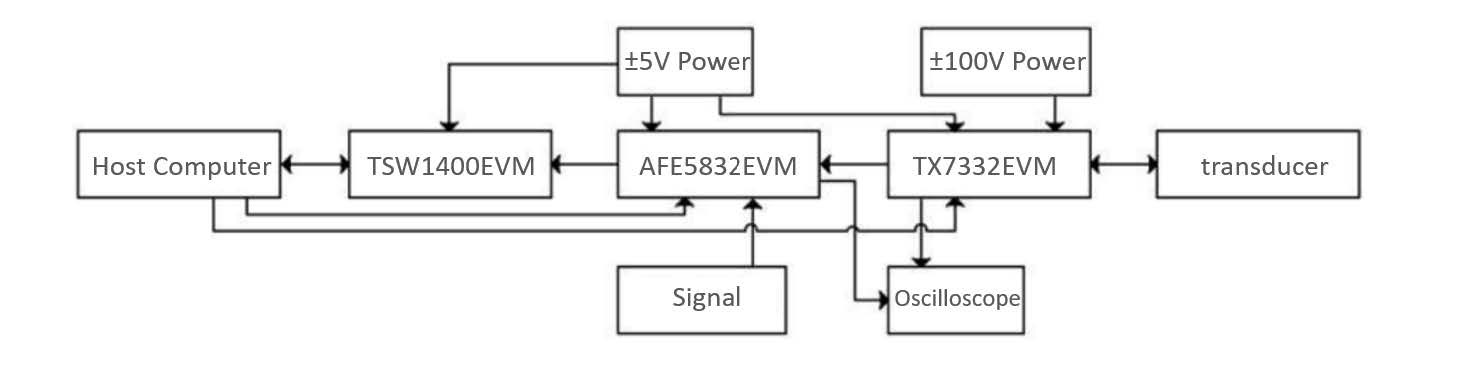}
	  \caption{Ultrasound platform system architecture}
      \label{FIG:1}
\end{figure*}

The TX7332 is a highly integrated, high-performance transmitter solution for ultrasound imaging system. The device has total 32 pulser circuits (PULS), 32 transmit/receive (T/R) switches, and supports both on-chip and off-chip beamformer (TxBF). The device also integrates on-chip floating power supplies that reduce the number of required high voltage power supplies.The TX7332 has a pulser circuit that generates three-level high voltage pulses (up to ±100 V) that can be used to excite multiple channels of an ultrasound transducer. The device supports total 32 outputs. The
maximum output current is configurable from 1.2 A to 0.3 A. A T/R switch under OFF state protects the receiver circuit by providing high isolation between the highvoltage transmitter and the low-voltage receiver when the pulser is generating high-voltage pulses. When the transducer is receiving echo signals, the T/R switch turns ON and connects the transducer to the receiver.\cite{11}

The AFE5832 device is a highly-integrated, analog front-end solution specifically designed for ultrasound systems where high performance, low power, and small size are required.The AFE5832 is an integrated analog front-end (AFE) optimized for medical ultrasound application. The device is realized through a multichip module (MCM) with three dies: two voltage-controlled amplifier
(VCA) dies and one analog-to-digital converter (ADC) die. Each VCA die has 16 channels and the ADC die converts all of the 32 channels. The ADC die has 16 physical ADCs. Each ADC converts two sets of outputs – one from each VCA die. The ADC is configured to operate with a resolution of 12 bits or 10 bits. The ADC resolution can be traded off with conversion rate, and operates at maximum speeds of 80 MSPS and 100 MSPS at 12-bit and 10-bit
resolution, respectively. The ADC is designed to scale its power with sampling rate. The output interface of the ADC comes out through a low-voltage differential signaling (LVDS) which can easily interface with lowcost field-programmable gate arrays (FPGAs).\cite{12}

The TSW1400EVM is a complete pattern generator and data capture board for evaluating most Texas Instruments high-speed analog-to-digital converters and digital-to-analog converters. The TSW1400EVM features a high-speed LVDS bus capable of providing 16-bit data at 1.5 GSPS. The board is equipped with 1GB of memory, providing 512 MB of 16-bit sampling depth.The TSW1400EVM series is available with an easy-to-use GUI software package. In pattern generation mode, HSDC (High Speed Data Converter) Pro can generate the required test patterns or load existing patterns for the DAC EVMs under test. The main features of the HSDC Pro software are: single-tone or multi-tone frequency testing, continuous data capture, channel power measurement, external trigger capability, master-slave operation, pattern generator, loading custom patterns, saving and exporting captured data, and frequency and time analysis.

\begin{figure}[h]
	\centering
		\includegraphics[scale=1]{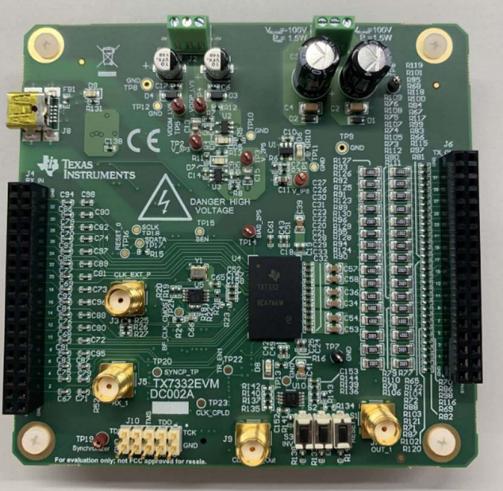}
	  \caption{TX7332EVM}
      \label{FIG:2}
\end{figure}

\begin{figure}[h]
	\centering
		\includegraphics[scale=1]{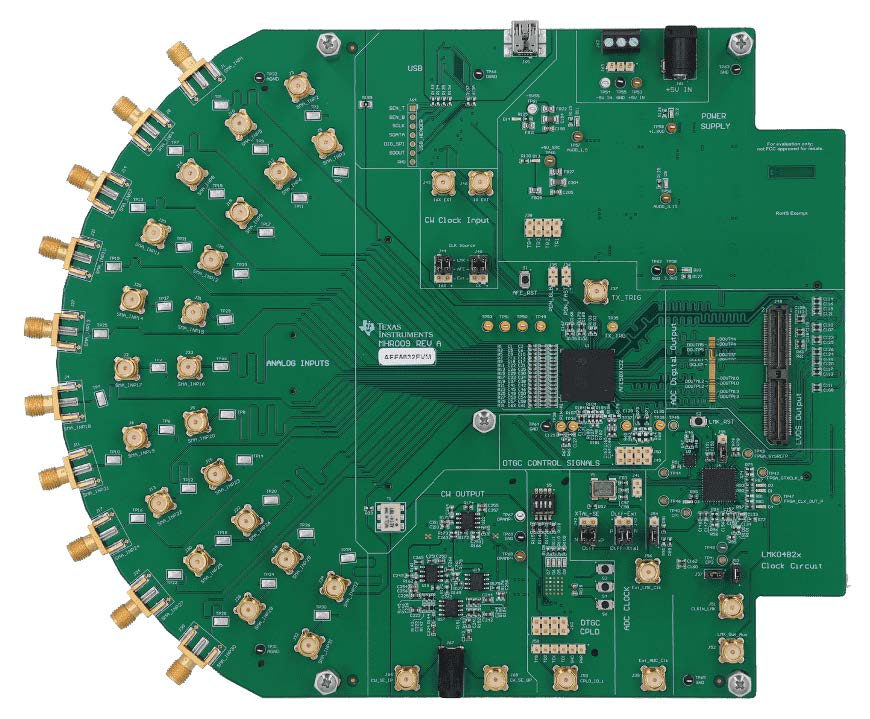}
	  \caption{AFE5832EVM}
      \label{FIG:3}
\end{figure}

The transducer in this product uses a customized 80-element linear array transducer (provided by
genosound, Shenzhen, China), which integrates acoustic elements such as matching layers, acoustic lenses,
and damping pads. The center frequency is 7MHz and the center distance of the array elements is 0.48mm.
We connected 32 array elements on one side of the transducer to build a verification platform.After the above boards are connected to the power supply and interconnected as shown in Figure 1, the
ultrasound imaging platform is assembled.After the platform is built, connect the positive and negative electrodes of a single probe with a center
frequency of 5MHz to any channel on the TX\_OUT interface on the TX7332EVM, and place the
probe in a stainless steel tank filled with water, so that the probe plane is parallel to the tank
wall to obtain the maximum echo energy. Control the TX7332EVM to transmit a signal with a frequency of
5MHz and a single 3-pulse signal. Observe the echo signal in the host computer software.

Connect the probe to the TX7332EVM, set the delay of all 32 channels to 0, and transmit a signal with a
frequency of 7MHz and 6 pulses per time. Immerse the top of the probe in a stainless steel tank filled with
water, and set the target to copper wire. After receiving the echo signal, output the RF data in .csv format perform software post-processing, and output the B-mode image.

\begin{figure}[h]
	\centering
		\includegraphics[scale=1]{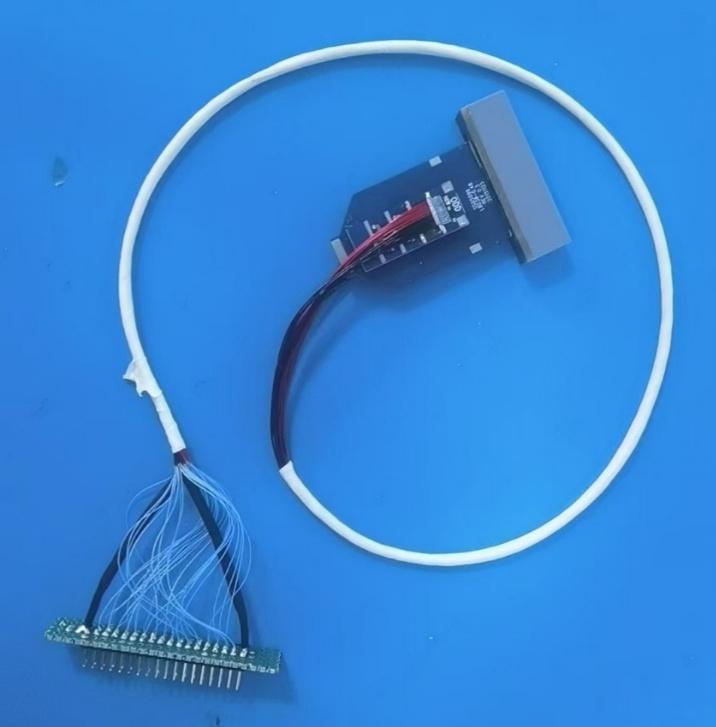}
	  \caption{Probe}
      \label{FIG:4}
\end{figure}

\begin{figure}[h]
	\centering
		\includegraphics[scale=0.6]{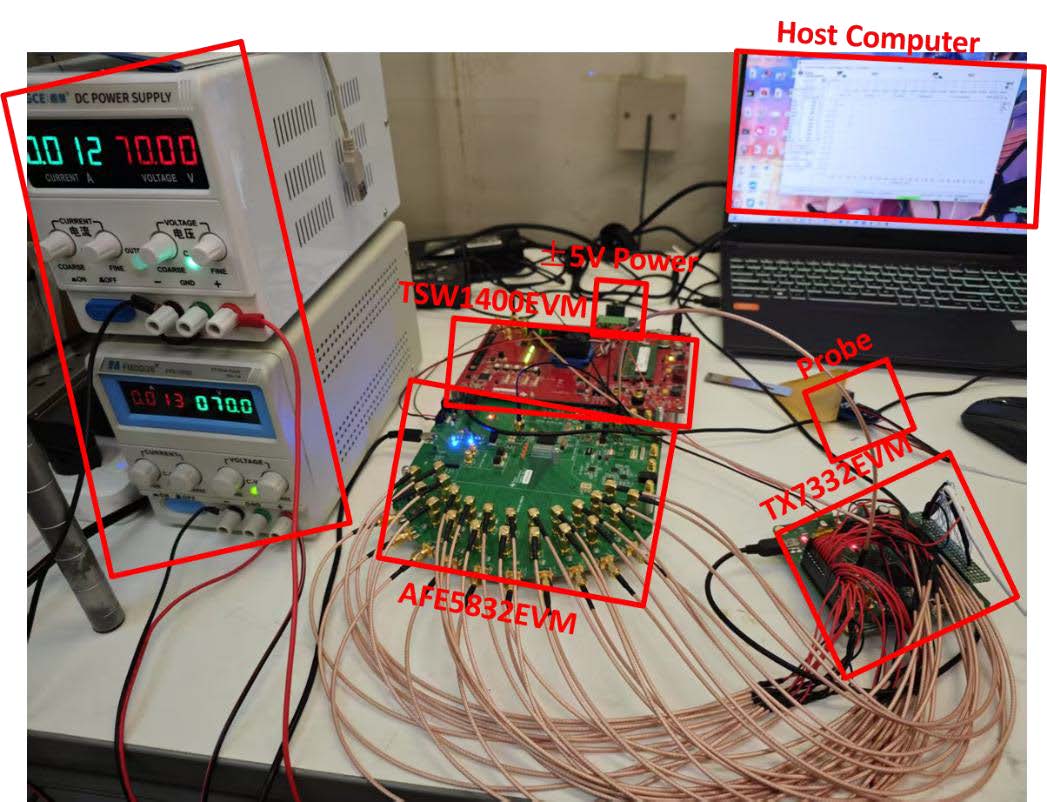}
	  \caption{Finished product}
      \label{FIG:5}
\end{figure}

\section{Design and fabrication of ultrasound linear array transducer}

A typical 1D array transducer is composed of an active layer, acoustic matching layers, a backing block, an acoustic lens, kerfs, a ground sheet (GRS), and a signal flexible printed circuit board (FPCB). The active layer is usually made of a piezoelectric material—mostly piezoceramic. The active layer generates an ultrasound wave in response to an electric driving signal, receives the wave reflected at the boundary of an organ, and converts the received ultrasound wave to an electric signal by means of the piezoelectric effect\cite{13}. However, the big difference in the acoustic impedance between piezoceramic elements and a human body prevents the efficient transfer of ultrasonic energy between the two media. So the job of a matching layer is to help transfer the ultrasound energy from the elements to the medium (flesh). The matching layer(s) are located between the elements and the lens. They are made of materials that are conducive to achieving better energy transfer, such as epoxy, polyurethane, polystyrene, etc. The backing layers prevent the backward emitted sound waves to echo and ring back into the transducer for detection.

When all the array elements of the phased array transducer are excited by the same signal at the same time, the main lobe sound beam generated is perpendicular to the array element surface, and the sound beam deflection angle is zero. If adjacent array elements are excited by the same excitation source at a certain time difference$\Delta t$, the sound pulses generated by each adjacent array element will also be delayed by $\Delta t$ accordingly. In this way, the main lobe of the synthesized beam is no longer perpendicular to the array, but forms an angle $\theta$ with the normal of the array, which is the deflection angle.The deflection angle $\theta$ is a function of $\Delta t$. If it remains $\Delta t$ unchanged and starts from n, and each array element chip is excited in turn, the main lobe beam will deflect to the corresponding direction on the opposite side of the array normal.\cite{14}According to the standing wave theory, the condition for the piezoelectric chip to resonate under the excitation of high-frequency electric pulses is:

$$
t=\lambda_L/2=C_L/2f_0
$$
Therefore, it is easy to calculate through formula 6 that the thickness of the piezoelectric ceramic sheet should be 0.295mm.

After the above calculations, we obtained the main parameters of the piezoelectric 2-2 composite material as shown in Table 1, and used the cut and fill process to obtain the finished product (provided by Xinwei electronic, BaoDin, China).

\begin{table*}[h]
\caption{main parameters of the piezoelectric 2-2 composite material}\label{tbl1}
\begin{tabular*}{\tblwidth}{lcccc}
\toprule
 Element Spacing/mm & Element Length/mm & Element thickness/mm & Element Width/mm & total length/mm \\ 
\midrule
0.17&10&0.295&0.14&10.88 \\
\bottomrule
\end{tabular*}
\end{table*}

To maximize energy transmission and minimize reflection, the matching layer is typically designed with a quarter-wavelength thickness ($\lambda$/4). According to transmission line theory, when a medium's thickness is one-quarter of the wavelength of the sound wave within it, it can effectively match two acoustically mismatched media. Specifically, the acoustic impedance of the matching layer should be the geometric mean of the piezoelectric material's and the biological tissue's acoustic impedances. Through this $\lambda$/4 thickness design, the matching layer generates reflected waves that are out of phase, thereby achieving maximum acoustic energy transmission at the transducer-tissue interface, significantly improving ultrasound signal sensitivity and penetration depth.Considering that the speed of sound in epoxy resin is 2450m/s, the design thickness of the matching layer should be 0.12mm.

In the first attempt, the scrape coating method was employed for fabricating the matching layer. The procedure involved dropping the prepared epoxy resin onto a glass substrate and then applying it with a scraper. However, this method encountered several issues: difficulty in demolding, uneven film thickness, and a significant number of bubbles within the film. These problems were attributed to inconsistent scraping speed, insufficient surface energy of the glass for proper epoxy spreading, and the lack of defoaming treatment for the epoxy resin prior to coating.

Following the shortcomings of the scrape coating method, the second attempt utilized spin coating. The process involved dropping the epoxy resin onto the glass substrate and then spin coating it at 1000 rpm for 2 minutes. Despite the change in method, challenges such as demolding difficulty and uneven film thickness persisted. Additionally, spin coating proved to be unsuitable for producing films of the desired thickness. These issues primarily stemmed from an improper selection of release agent and the continued insufficient surface energy of the glass for adequate epoxy spreading.

The final attempt involved a pressing method. The procedure was as follows: after preparing the epoxy resin, it was dropped onto a glass substrate that has been pre-coated with water repellent (rain-x, ITW Global Brands, Houston, USA) and two alumina flakes with a thickness of 0.12 mm were placed on the glass to act as spacers. Another piece of glass was then used to press the assembly. This method aimed to control the film thickness and improve uniformity through physical compression.\cite{15}

Thanks to Darmawan’s research\cite{16}, we know that composites prepared using 1 $\mu$m monodisperse metal particles and a 4:1 weight ratio show acceptable functionality as an ultrasonic transducer backing layer. We mixed tungsten powder and epoxy resin at a mass ratio of 4:1 and stirred them evenly. Then, we placed them in a vacuum environment and continued to stir for 10 minutes to defoam and evenly disperse the tungsten powder particles. Then, we poured them into a PTFE mold and placed them in a constant temperature drying oven at 120°C for 2 hours to cure. After cooling and demolding, we got a backing layer with suitable acoustic properties.

Due to the supplier's process limitations for the 2-2 composite material, it was impossible to retain independent electrodes after processing. Therefore, the engineering problem we faced was to develop a method to plate independent electrodes for each array element ourselves.To address this problem, the first step involved fabricating a high-precision mask and a fixing sheet. The mask was created by laser-cutting a stainless steel sheet, ensuring that the gap spacing of the mask precisely matched the element spacing of the 2-2 composite material to guarantee accurate plating patterns. Concurrently, a hollow stainless steel sheet of the same thickness as the composite material was processed to serve as a holder for fixing the composite, with its precise thickness being critical for stability during the plating process.

Prior to plating, all materials, including the slides, fixing sheet, mask, and composite material, had to be thoroughly cleaned with alcohol to ensure their surfaces were free of grease and dust, a critical step in preventing plating defects. Subsequently, double-sided tape with varying adhesive strengths was used to firmly attach the fixing sheet to the slide, with a portion of the tape intentionally exposed at the top to provide additional adhesion, preventing lateral movement of the composite during operation. Initial challenges might include insufficient surface cleanliness or insecure fixation, requiring careful attention.

Finally, magnetron sputtering was performed. The composite material was covered with the mask and precisely aligned under a microscope, ensuring the mask gaps aligned with the composite material's elements. Once secured, it was placed into the sputter machine, and the process was initiated (parameters set to 30mA current for 120 seconds). Due to the limitations of mask processing technology, an excessively wide mask would cause severe stainless steel deformation. Consequently, one piece of composite material could not be fully plated in a single run, necessitating two separate sputtering steps. The greatest challenge here was ensuring that the electrodes plated in these two separate runs were seamlessly connected to guarantee conductivity and prevent issues like "no connection between top and bottom." Precise alignment under the microscope was crucial for correct electrode pattern formation and achieving good connectivity between the two plating stages.

\begin{figure}[h]
	\centering
		\includegraphics[scale=0.4]{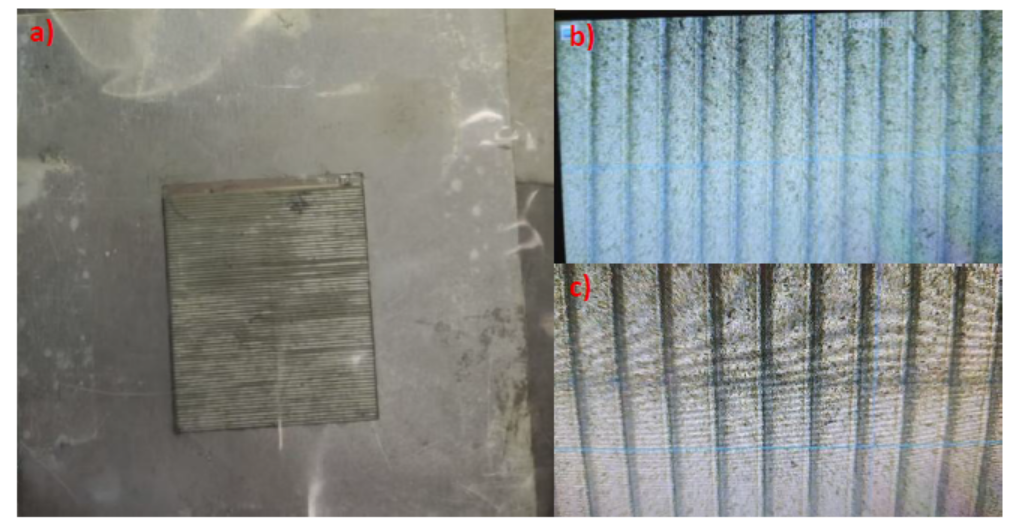}
	  \caption{a): The surface is dirty; b): Not fixed firmly; c): No connection between top and bottom}
      \label{FIG:6}
\end{figure}

\begin{figure}[h]
	\centering
		\includegraphics[scale=0.4]{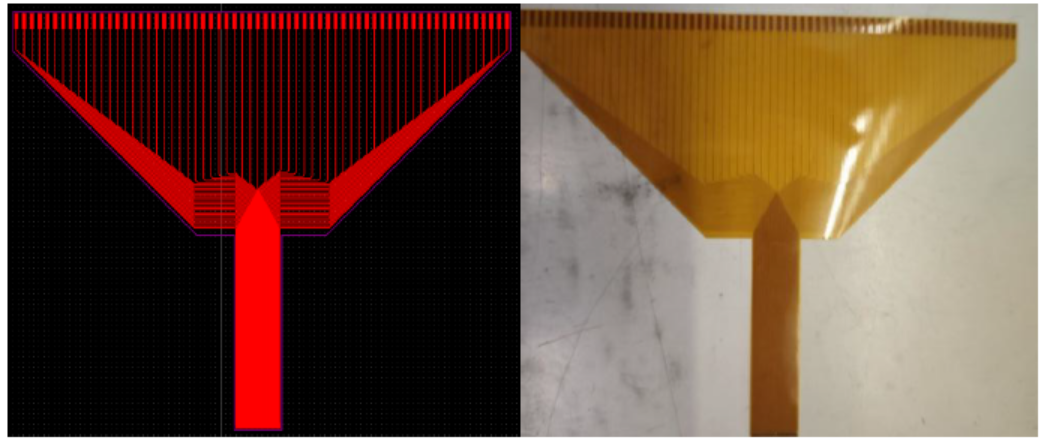}
	  \caption{FPC design and finished product}
      \label{FIG:7}
\end{figure}

\begin{figure*}[h]
	\centering
		\includegraphics[scale=1]{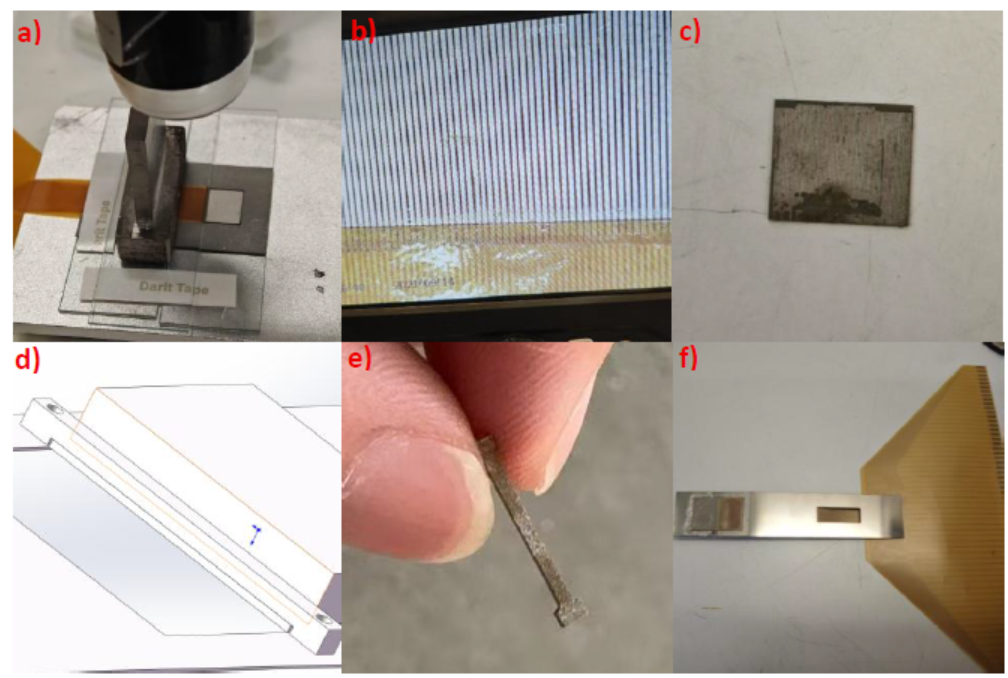}
	  \caption{a),b): epoxy and pressing; c): silver electrodes peel off; d),e): structural parts for physical pressing; f): semi-finished}
      \label{FIG:8}
\end{figure*}

Since the array element width of the transducer was too thin to directly weld copper wire, a Flexible Printed Circuit (FPC) was designed to connect the transducer and external cables. However, the connection between the transducer array electrodes and the FPC electrodes became a significant problem.The first attempt involved bonding with epoxy and applying pressure during curing. The FPC was aligned with the transducer electrodes under a microscope, then bonded with epoxy and cured under pressure. However, this method was not ideal, as approximately half of the electrodes failed to conduct. This indicated that simple epoxy bonding and pressure were insufficient to ensure reliable connections for all the tiny electrodes.The second attempt aimed to pre-tin the FPC electrodes with solder paste and then use a hot press to solder them to the transducer electrodes. This method sought to establish a more reliable electrical connection through soldering. However, this attempt also encountered problems: the tin balls in the solder paste were too large, making bridging (short circuits between adjacent electrodes) highly probable. Even if the bridging issue was resolved, the silver electrodes themselves had low adhesion and tended to peel off easily after heating, resulting in unstable connections.

The third attempt considered using structural parts for physical pressing. This method aimed to ensure intimate contact between the transducer electrodes and FPC electrodes through mechanical pressure, thereby achieving electrical conductivity. Unfortunately, during the project, the fabricated transducer and remaining raw materials were destroyed in an accident, preventing the opportunity to verify the feasibility of this solution. Ultimately, to demonstrate a semi-finished product, tape was used to temporarily secure it, but further reliability tests could not be conducted as it had not yet been waterproofed.

\section{Summary}

This report details the successful construction of an ultrasound imaging platform and the design and fabrication of a novel ultrasound endoscope probe. The primary objective of this project was to establish a functional system for acquiring and processing ultrasound signals, specifically targeting minimally invasive endoscopic applications.

Regarding the ultrasound imaging platform, it was designed and developed based on Texas Instruments (TI) Evaluation Modules (EVMs). This platform is capable of transmitting 32-channel high-voltage signals and receiving echo signals, featuring on-chip signal amplification and acquisition capabilities. Furthermore, the platform integrates a complete Time Gain Control (TGC) imaging path and a Continuous Wave Doppler (CWD) path, supporting imaging with linear array, convex array, and phased array probes in conjunction with host computer software.

In terms of the design and fabrication of the ultrasound endoscope probe, a 64-element, 5MHz center frequency, phased array linear ultrasound endoscopic probe was designed, aiming for miniaturization and optimal imaging performance. The report meticulously documented the various attempts and challenges encountered during the fabrication and assembly of its matching layer, backing layer, 2-2 piezoelectric composite material, and electrodes. Although difficulties were faced in areas such as FPC connection, and the final proposed solution could not be fully verified due to an unforeseen accident, various connection methods were explored and their respective advantages and disadvantages analyzed.

Despite numerous engineering challenges, this project successfully established a functional ultrasound imaging system and completed the preliminary design and fabrication of its critical component: a miniaturized ultrasound endoscope probe. Future work will focus on further optimizing the probe's manufacturing process, particularly addressing the reliable connection between the FPC and array electrodes, and conducting comprehensive performance testing and waterproofing to realize its full potential in practical minimally invasive endoscopic applications.










\bibliographystyle{cas-model2-names}

\bibliography{cas-refs}



\end{document}